\documentclass[sigconf]{aamas}

\usepackage{balance}
\usepackage{url}
\usepackage{xcolor}
\definecolor{newcolor}{rgb}{.8,.349,.1}
\usepackage{siunitx}

\usepackage{amsmath}
\usepackage{mathtools}
\usepackage{enumitem}
\usepackage{wrapfig}

\definecolor{darkred}{rgb}{0.6,0,0}
\definecolor{lightgreen}{rgb}{0.1,0.8,0.1}
\definecolor{darkgreen}{rgb}{0.1,0.3,0.1}
\definecolor{middlegreen}{rgb}{0.1,0.5,0.1}
\definecolor{darkorange}{rgb}{0.6,0.4,0.2}
\definecolor{darkPurple}{rgb}{0.4,0.1,0.4}
\definecolor{lightblue}{rgb}{0.5,0.5,0.75}
\definecolor{blue}{rgb}{0,0,0.75}
\definecolor{orchid}{rgb}{0.5,0,0.5}
\definecolor{orange}{rgb}{1.0,0.3,0}
\definecolor{darkorange}{rgb}{0.6,0.2,0}

\newcommand{\commentnew}[2]{\textcolor{#1}{#2}}

\newcommand{\yf}[1]{\commentnew{black}{  #1}}

\AtBeginDocument{%
  \providecommand\BibTeX{{%
    \normalfont B\kern-0.5em{\scshape i\kern-0.25em b}\kern-0.8em\TeX}}}

\setcopyright{ifaamas}
\acmConference[AAMAS '21]{Proc.\@ of the 20th International Conference on Autonomous Agents and Multiagent Systems (AAMAS 2021)}{May 3--7, 2021}{Online}{U.~Endriss, A.~Now\'{e}, F.~Dignum, A.~Lomuscio (eds.)}
\copyrightyear{2021}
\acmYear{2021}
\acmDOI{}
\acmPrice{}
\acmISBN{}

\begin{document}

\title{It's A Match! Gesture Generation Using Expressive Parameter Matching} 
\subtitle{Extended Abstract}

\author{Ylva Ferstl}
\email{yferstl@tcd.ie}
\affiliation{%
  \institution{Trinity College Dublin}
  \streetaddress{College Green, Dublin 2}
  \city{Dublin}
  \country{Ireland}
}

\author{Michael Neff}
\email{mpneff@ucdavis.edu}
\affiliation{%
  \institution{University of California Davis}
  \streetaddress{1 Shields Ave}
  \city{Davis}
  \state{California}}

\author{Rachel McDonnell}
\email{ramcdonn@tcd.ie}
\affiliation{%
  \institution{Trinity College Dublin}
  \streetaddress{College Green, Dublin 2}
  \city{Dublin}
  \country{Ireland}
}

\begin{abstract}

\yf{Automatic gesture generation from speech generally relies on implicit modelling of the nondeterministic speech-gesture relationship and can result in averaged motion lacking defined form. Here, we propose a database-driven approach of selecting gestures based on specific motion characteristics that have been shown to be associated with the speech audio.}
\yf{We extend previous work that identified  expressive parameters of gesture motion that can both be predicted from speech and are perceptually important for a good speech-gesture match, such as gesture velocity and finger extension.}
A perceptual study was performed to evaluate the appropriateness of the gestures selected with our method. We compare our method with two baseline selection methods. The first respects timing, the desired onset and duration of a gesture, but does not match gesture form in other ways. The second baseline additionally disregards the original gesture timing for selecting gestures. The gesture sequences from our method were rated as a significantly better match to the speech than gestures selected by either baseline method.

\end{abstract}



\maketitle

\section{Introduction}

\yf{Much prior work has explored methods for automating generation of gesture behavior from speech (e.g. \cite{ferstl2019multi,kucherenko2020genea,ondras2020audio}) or text input (e.g. \cite{yoon2019robots}). An advantage of models based on speech audio is the readily obtainable input signal, as well as fast and automatic prosody processing software. Unfortunately, machine learning-based gesture generation models often produce averaged motion that lacks realistic and defined gesture form due to training models on exact joint positions or angles and failing to capture the nondeterministic relationship between gesture and speech. Here, we propose the use of five expressive motion parameters that have been shown in previous work to be associated with the speech signal as criteria for selecting gestures matched to the speech audio. We combine machine learning and database sampling by first estimating gesture parameters from speech prosody using machine learning techniques, and then searching a large gesture database for gestures that best match the predicted expressive parameters.} We evaluate the suitability of this method of gesture selection with a perceptual study in an ablation manner.

\section{Gesture matching method}\label{algorithm}
A key component of our method is a large dataset of individual gestures, allowing production of a large variety of gesture behavior without requiring a generative model. We use 10 hours of data combined from two datasets of conversational data, each with one speaker (\cite{ferstl2019multi} and \cite{ferstl2018investigating}). The motion data was segmented by labelling the stroke phases, the expressive phase of a gesture \cite{kendon1972some} by using the method of \cite{ferstl2020adversarial}, resulting in a total of almost 23,700 gesture strokes. 
Each stroke is automatically labelled with its gesture parameter values through motion analysis. Briefly, for a new speech utterance, we compute the desired gesture parameters from the speech prosody; then, we search the gesture database for the best match with respect to the parameter values and join gestures into sequences by synthesizing preparation and retraction phase for the selected gesture. For estimating the five expressive parameters of a gesture from a speech signal, we rely on the method of \cite{ferstl2020understanding}.

\yf{The system inputs are the gestures' timings as well as the associated speech segments. Gesture timings are defined by the annotated stroke timings. We then determine the best matching gesture for each stroke slot by estimating five gesture parameters from the associated speech audio:} (1) The gesture velocity, (2) the size of the initial acceleration peak, (3) the gesture size measured by the total completed path length, (4) the arm swivel (the rotation around the axis between the shoulder and the wrist, bringing the elbow closer or further from the body), and (5) hand opening, describing how open or closed the hand shape is (calculated as the mean distance of the finger tips from the base of the wrist).
\yf{Previous work has described how these five gesture parameters can be estimated from speech, and shown that they affect perceptions of how well the gestures match the speech \cite{ferstl2020understanding}.}

Given the computed gesture parameters for a speech segment, we search the database for the best match. First, each gesture in the database is assigned a rank number with respect to each of the five parameters; e.g. the gesture with the closest-matching velocity would receive velocity rank 1, and the gesture sample with the worst-matching velocity receives velocity rank 23,700. Each gesture hence has 5 rank values, one for each parameter. 
Each rank value is weighted to decide the importance of a parameter, before summing all 5 rank values into a total match rank. The gesture with the best (smallest) match rank is selected. 
As we use pre-determined stroke durations from the labelled stroke phase input, we also constrain gesture selection to gestures with similar duration. This ensures that a selected gesture will not overlap with the next stroke slot and hence alter the gestures' timings. 
The selected gestures are combined into coherent sequences using animation software based on the open-source animation environment DANCE \cite{shapiro2005dynamic}, taking as input motion data and corresponding stroke labels and synthesizing transitions between gestures and to a rest pose using motion interpolation.

\section{Evaluation}
To evaluate how well the gesture selection matches the speech expression, we performed a perceptual experiment comparing our method in an ablation manner to two baseline methods, as well as to the ground truth gestures. 
Our first baseline comparison uses the same stroke timing but selects gestures agnostic to the desired parameter values, i.e. the first baseline method (unmatched) is equivalent to our method without calculating the match rank. 
The {\em baseline 1} comparison gives indication of the importance of matching expressive gesture parameters to the speech.

The second baseline method does not use the same stroke timing (unmatched \& untimed); it scrambles the order of all the timings within our test dataset, resulting in the same realistic stroke and between-stroke durations, without preserving the speech-gesture synchrony. The {\em baseline 2} comparison assesses the importance of correct gesture timing.

The ground truth condition selects the true stroke for each gesture slot. This therefore reflects the true gesture behavior, but uses synthetic preparations and retractions.

There were 48 experiment trials (6 clips for each of the 2 speakers, for 4 conditions), where each clip contained a different speech segment. We first ran a pre-test on the ground-truth gesture version of the 48 speech segments with 55 participants (21 females, 33 males, 1 other, ages - years, \textit{M} = 22.0, \textit{SD} = 6.6) where we verified that variations between selected segments were not caused by variance in gesture timing.
Subsequently, 54 participants completed our online experiment (12 females, 41 males, 1 other, ages - years, \textit{M} = 20.8, \textit{SD} = 4.3) where trials were presented in random order.

We visualize the generated gesture sequences on the Brad character from the open-source Virtual Human Toolkit (VHTK) \cite{hartholt2013all}. Participants rated gesture sequences on a 7-point Likert scale with respect to the following question:
\textit{``How well did the expressive quality of the gestures match 
the expressive quality of the speech?''}
(All stimuli: \url{https://www.youtube.com/playlist?list=PLLrShDUC_FZwrW0Oc9GTTdrBeMGpsA4PB})

\begin{wrapfigure}{R}{0.25\textwidth} 
\vspace{-10pt}
  \begin{center}
    \includegraphics[width=0.25\textwidth]{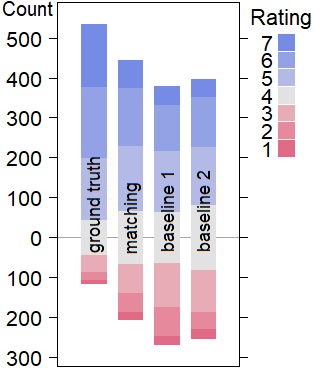}
    \caption{Stacked bar chart of perceptual ratings. Plotted is the frequency of responses for each of the 7 rating scores. (The y-axis is the frequency of responses)}
    \label{perceptualResults}
  \end{center}
  \vspace{-10pt}
\end{wrapfigure} 

Statistical analysis of the results of the perceptual experiment was performed by treating the Likert rating scores as ordinal data and fitting a cumulative link model, using clm from the R ordinal package \cite{christensen2015ordinal}.
The ground truth condition was rated significantly higher than all other conditions (all $p<.001$), indicating that the ground truth gestures were preferred over any alternative, as expected (rating score $mean= 5.35$). The gesture matching condition (our method) was rated significantly higher ($mean = 4.67$) than both baseline methods (both $p<.001$, baseline 1 $mean = 4.32$, baseline 2 $mean = 4.41$), indicating that matching gestures to speech-predicted parameters indeed increases perceived appropriateness of the selected gesture, in line with our hypothesis.
The two baseline conditions were not rated significantly different from each other, suggesting, somewhat surprisingly, that correct gesture timing alone did not improve perceived speech-gesture match. Both speakers have a relatively high gesture frequency; for speakers with less frequent gestures, wrong gesture timing may be more perceivable. Results are visualized in Figure \ref{perceptualResults}.

\section{Conclusion}

Here, we evaluated the use of matching expressive gesture parameters to speech for generating gesture behavior for conversing virtual characters.
Our results show that our gesture selection method does indeed perform better than the baselines disregarding parameter matches, indicating that matching the chosen gesture parameters to the speech improves perceived speech-gesture match. This confirms the proposition of \shortcite{ferstl2020understanding} on the importance of matching expressive gesture parameters to co-occurring speech, and asserts the validity of the here proposed avenue for gesture generation. 

Interestingly, baseline 2, which does not use any timing or speech prosody information, still received relatively high ratings and did not differ significantly from baseline 1, which used correct timing. One potential reason for this is that the speakers in the used datasets produce continuous speech without any significant periods of silence, therefore even untimed gesture is almost always accompanied by speech. In cases without continuous speech, untimed gestures may stand out more negatively. 
Alternatively, realistic gesture forms could be enough for reasonably well-liked gesture performances. 
Notably, \shortcite{fernandez2014gesture} have also reported greater importance of matching gesture quality (measured by stroke intensity) than matching speech timing.

In this study, we evaluate our gesture selection by using the gesture stroke locations determined by motion segmentation which is not feasible in generative systems as it requires access to the ground-truth motion. In future work, however, we plan to combine our gesture selection with methods for determining stroke timing that only rely on speech audio in order to create a fully speech-based system. 
\footnote{This research was funded by Science Foundation Ireland under the ADAPT Centre for Digital Content Technology (Grant 13/RC/2106).}

\balance
\bibliographystyle{ACM-Reference-Format}
\bibliography{bibliography}

\end{document}